\documentclass[aps,pra,twocolumn,floatfix,showpacs]{revtex4}
\usepackage[dvips]{color}
\usepackage{amsmath,amssymb,amsfonts,bm,graphicx,latexsym}
\usepackage[hypertex,breaklinks=true,colorlinks=false]{hyperref}
\newcommand{\cutspace}{}

\newcommand{\ua}{\uparrow}
\newcommand{\da}{\downarrow}

\newcommand{\Scal}{{\cal S}}
\newcommand{\group}{\mathrm{group}}
\newcommand{\ThD}{\mathrm{(3D)}}

\newcommand{\Kv}{\bm{K}}

\newcommand{\Psit}{\tilde{\Psi}}

\newcommand{\Nb}{ {\bar{N}} }

\begin{document}

%%%%%%%%%%%%%%%%%%%%%%%%%%%%%%%%%%%%%%%%%%%%%%%%%
% Paper Information
%%%%%%%%%%%%%%%%%%%%%%%%%%%%%%%%%%%%%%%%%%%%%%%%%
\title{
Quantum Hall States in Rapidly Rotating Two-Component Bose Gases
}

\author{Shunsuke Furukawa}
\affiliation{Department of Physics, University of Tokyo, 7-3-1 Hongo, Bunkyo-ku, Tokyo 113-0033, Japan}

\author{Masahito Ueda}
\affiliation{Department of Physics, University of Tokyo, 7-3-1 Hongo, Bunkyo-ku, Tokyo 113-0033, Japan}
%\affiliation{Macroscopic Quantum Control Project, ERATO, JST, Bunkyo-ku, Tokyo 113-8656, Japan}

\date{\today}
%\keywords{}

% 03.75.Hh Static properties of condensates; thermodynamical, statistical, and structural properties
% 03.75.Lm Tunneling, Josephson effect, Bose-Einstein condensates in periodic potentials, solitons, vortices, and topological excitations 
% (see also 74.50.+r Tunneling phenomena; Josephson effects in superconductivity)
% 03.75.Mn Multicomponent condensates; spinor condensates
% 05.30.Jp Boson systems 
% (for static and dynamic properties of Bose-Einstein condensates, see 03.75.Hh and 03.75.Kk; see also 67.10.Ba Boson degeneracy in quantum fluids)
% 73.43.-f Quantum Hall effects
% 73.43.Cd Theory and modeling
% 73.43.Nq Quantum phase transitions

%%%%%%%%%%%%%%%%%%%%%%%%%%%%%%%%%%%%%%%%%%%%%%%%%
% Abstract
%%%%%%%%%%%%%%%%%%%%%%%%%%%%%%%%%%%%%%%%%%%%%%%%%
\begin{abstract}
We investigate strongly correlated phases of two-component (or pseudo-spin-$1/2$) Bose gases under rapid rotation 
through exact diagonalization on a torus geometry. 
In the case of pseudo-spin-independent contact interactions, 
we find the formation of gapped spin-singlet states 
at the filling factors $\nu=k/3+k/3$ ($k/3$ filling for each component) with integer $k$. 
We present numerical evidences that 
the gapped state with $k=2$ is well described as a non-Abelian spin-singlet (NASS) state, 
in which excitations feature non-Abelian statistics. 
Furthermore, we find the phase transition from the product of composite fermion states 
to the NASS state by changing the ratio of the intercomponent to intracomponent interactions. 
\end{abstract}
\maketitle

%%%%%%%%%%%%%%%%%%%%%%%%%%%%%%%%%%%%%%%%%%%%%%%%%
% Main text
%%%%%%%%%%%%%%%%%%%%%%%%%%%%%%%%%%%%%%%%%%%%%%%%%

%--------------------------------------
% [ Introduction ]

Under rapid rotation, ultracold gases of bosonic atoms are predicted to enter a highly correlated regime, 
analogous to quantum Hall systems \cite{Cooper08_review, Wilkin98}. 
This regime is reached as the number of vortices, $N_V$, in a Bose-Einstein condensate 
becomes comparable with the number of atoms, $N$. 
The relevant control parameter is the filling factor $\nu=N/N_V$. 
For scalar bosons, it is predicted that the vortex lattice melts for $\nu\lesssim 6$ \cite{Cooper01}
and that a series of gapped uncondensed states appear at various integer and fractional $\nu$. 
In particular, the ground states at $\nu=k/2$ (with $k=1,2,3,...$) have large overlaps \cite{Cooper01,Regnault07} with the Read-Rezayi states \cite{Read99}, 
whose excitations feature non-Abelian statistics for $k\ge 2$. 
Other quantum Hall states such as composite fermion states \cite{Regnault03} have also been discussed. 

%--------------------------------------
% [ Our paper ]

Given a rich diversity of strongly correlated physics in the scalar case, 
it is natural to ask what happens in two-component Bose gases, 
such as those made up of two hyperfine spin states of the same atoms \cite{Hall98,Schweikhard04}. 
For intermediate rotation frequencies, 
a variety of vortex lattices have been shown to appear \cite{Mueller02, Kasamatsu03}. 
For a rapid rotation, 
two-component systems would offer an ideal situation 
in which to study the roles of (pseudo-)spin degrees of freedom in the quantum Hall physics. 
As the ratio of the intercomponent contact interaction $g_{\ua\da}$ to the intracomponent one $g$ increases, 
the two spin states are expected to be entangled to form novel ground states. 

On the experimental front, the smallest filling factor achieved for rotating scalar bosons 
is $\nu \approx 500$ \cite{Schweikhard04_scalar}, which is far above the quantum Hall regime. 
The main difficulty in rotating a trapped gas faster is a fine tuning of the rotation frequency without passing through the deconfinement limit, 
and novel ideas for circumventing this problem using anharmonic traps have been proposed \cite{Morris07,Roncaglia11}. 
In order to create a rapidly rotating two-component gas, 
one can first rotate a one-component gas and then convert a portion of atoms into another component 
by applying pulse lasers \cite{Hall98,Schweikhard04}. 
Furthermore, a very recent realization of an optically synthesized gauge field \cite{Lin09} has opened up a new route to smaller $\nu$. 
With these developments, 
there is now a growing prospect for realizing the quantum Hall regime of scalar or two-component bosons in the foreseeable future. 

In this Rapid Communication, we investigate strongly correlated phases 
of two-component Bose gases under rapid rotation 
by using the exact diagonalization method. 
For pseudo-spin-independent interactions $g_{\ua\da}=g$, 
we find the formation of gapped uncondensed ground states 
at filling factors $\nu=k/3+k/3$ ($k/3$ filling for each component) with integer $k$. 
Precisely at these filling factors, Ardonne and Schoutens \cite{Ardonne99} proposed 
a series of non-Abelian spin-singlet states (NASS) having a global $SU(3)_k$ symmetry; 
these states may be viewed as a spin-singlet generalization of Read-Rezayi [$SU(2)_k$] states 
and host non-Abelian anyonic excitations \cite{Ardonne99,Ardonne01}.  
We present numerical evidences that the gapped state with $k=2$ is well described as the $SU(3)_2$ state. 
We also discuss a phase transition that occurs at a particular value of the coupling ratio $g_{\ua\da} / g$. 
We note that NASS states of bosons have also been discussed 
in spin-$1$ bosons \cite{Reijnders02} 
and scalar bosons on optical lattices \cite{Hormozi12}. 
% We comment that with a similar spirit, strongly correlated phases of 
% spin-$1$ Bose gases have been studied in Ref.~\cite{Reijnders02}, 
% where different kinds of NASS states have been proposed. 
The two-component case studied here offers the simplest realistic setting 
which allows a detailed numerical identification of a NASS state. 

%--------------------------------------
% [ System setting ]

We consider a system of a Bose gas 
having two spin states (labeled by $\alpha=\ua,\da$) and 
rapidly rotating in a harmonic trap with the cylindrical symmetry about the $z$ axis. 
We denote the axial and radial trap frequencies as $\omega_\parallel$ and $\omega_\perp$, respectively. 
In the rotating frame of reference, 
a system of neutral particles is mathematically equivalent to that of charged particles 
subject to a uniform magnetic field. 
In analogy with the quantum Hall problem, we introduce the ``magnetic'' length $\ell=\sqrt{\hbar/(2M\Omega)}$, 
where $M$ is the particle's mass and $\Omega$ is the rotation frequency. 

%--------------------------------------
% [ Rapid rotation limit ]

As $\Omega$ increases toward $\omega_\parallel$, 
the particle density decreases due to the centrifugal spreading of the gas. 
For $\Omega\approx \omega_\parallel$, 
we may assume that the mean interaction energy per particle 
(roughly proportional to the coupling constants times the particle density) 
is much smaller than single-particle energy-level spacings, $\hbar \omega_{\parallel,\perp}$. 
The single-particle states can then be restricted to the ground state of the axial confinement 
and to the lowest Landau level (LLL) in the transverse directions \cite{Cooper08_review,Wilkin98}. 
The LLL state with angular momentum $L_z=m \hbar \ge 0$ in the $xy$ plane 
is represented as $u_m(z)\propto z^m \exp [ -|z|^2/(4\ell^2) ]$ with $z=x+iy$. 
Within this restricted subspace, 
we consider the interaction Hamiltonian consisting of intracomponent and intercomponent contact interactions: 
\begin{equation}\label{eq:Hint}
 H_\mathrm{int} 
 = \sum_{\alpha=\ua,\da} g_\alpha \sum_{i<j}^{N_\alpha} \delta (z_i^\alpha-z_j^\alpha) 
 + g_{\ua\da}  \sum_{i=1}^{N_\ua} \sum_{j=1}^{N_\da} \delta (z_i^\ua-z_j^\da) , 
\end{equation}
where $N_\alpha$ is the number of particles in the state $\alpha$ 
and $z_i^\alpha$'s are the positions of such particles. 
The effective coupling constants in the two-dimensional plane are given by
$g_\alpha = a_\alpha \sqrt{8\pi\hbar^3 \omega_\parallel / M}$ and 
$g_{\ua\da} = a_{\ua\da} \sqrt{8\pi\hbar^3 \omega_\parallel / M}$ \cite{Comment_contact}, 
where $a_\alpha$ and $a_{\ua\da}$ are the $s$-wave scattering lengths between like and unlike bosons, respectively. 
For simplicity, we assume $g_\ua=g_\da (\equiv g)$ in the following. 

%--------------------------------------
% [ Torus geometry ]

To study bulk properties, 
it is useful to work on closed uniform surfaces such as a sphere and a torus, 
which have no edge. 
Here, we perform calculations on a periodic rectangular geometry (a torus) \cite{Yoshioka84,Cooper08_review} 
of sides $L_x$ and $L_y$, 
which contains $N_V=L_x L_y /(2\pi \ell^2)$ vortices for each component.  
This geometry can describe the central region of a trapped atomic gas where the particle density is approximately uniform. 
There are $N_V$ LLL states for each component. 
We introduce the filling factor as $\nu=N/N_V$, 
where $N=N_\ua+N_\da$ denotes the total particle number. 
We classify all states by the pseudomomentum
$\Kv=(K_x,K_y)=2\pi\hbar (m_x/L_x, m_y/L_y)$ with integer $m_x$ and $m_y$. 
At $\nu= p/q$ with $p$ and $q$ being coprime, 
all energy eigenstates possess a trivial center-of-mass degeneracy of $q$, 
and the minimal Brillouin zone (without this degeneracy) consists of $\Nb^2$ points, 
where $\Nb$ is the largest common divisor of $N$ and $N_V$ \cite{Haldane85}.  
As in Ref.~\cite{Cooper01}, we focus on non-negative values of $K_x$ and $K_y$ up to the Brillouin zone boundary, 
since states at $(\pm K_x, \pm K_y)$ are degenerate by symmetry. 
Henceforth, we take the units in which $\hbar\equiv 1$ and $\ell\equiv 1$. 

%############################
\begin{figure}
\begin{center}
\includegraphics[width=0.42\textwidth]{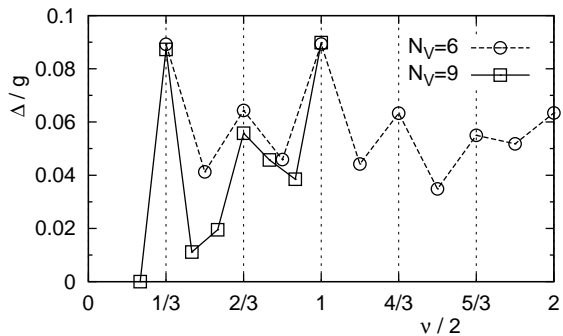}
\cutspace
\end{center}
\caption{
``Charge'' gap $\Delta (N)$ [Eq.~\eqref{eq:charge_gap}] 
versus the filling factor per component, $\nu/2=N/(2N_V)$, in the $SU(2)$-symmetric case $g_{\ua\da}=g$. 
The aspect ratio is set to $L_x/L_y=3/(2\sqrt{3})$ and $3/(3\sqrt{3})$ for $N_V=6$ and $9$, respectively. 
These ratios are chosen to be compatible with the rectangular vortex lattice expected for large $\nu$ \cite{Mueller02}. 
}
\label{fig:Gap}
\end{figure}
%############################

%--------------------------------------
% [ SU(2) case ]

We first focus on the case of pseudo-spin-independent interactions $g_{\ua\da}=g$, 
where the system possesses an $SU(2)$ symmetry. 
For this case, a mean field theory \cite{Mueller02} predicted 
that each component forms a rectangular vortex lattice with the aspect ratio of the unit cell given by $\sqrt{3}$. 
In analogy with the scalar case \cite{Cooper01}, 
we expect to find a phase transition from the vortex lattice 
to gapped uncondensed states as the filling factor $\nu$ decreases. 
We therefore set the aspect ratio of the torus to be compatible with this lattice, 
such that this lattice can be detected if it appears. 
For $N_V=6$, for example, we set $L_x/L_y = 3/(2\sqrt{3})$, 
with which the system can host $3\times 2$ units of the rectangular vortex lattice, 
with the longer side of the rectangular unit in the $y$ direction. 

%--------------------------------------
% [ ``Charge'' gap ]

To search for uncondensed ground states, 
we introduce the ``charge'' gap
\begin{equation}\label{eq:charge_gap}
\begin{split}
 \Delta (N) 
%   = & E\left( N/2+1, N/2 \right) + E\left( N/2-1, N/2 \right) \\
%   &- 2 E\left( N/2, N/2 \right), 
 = & E\left( \frac{N}{2}+1, \frac{N}{2} \right) + E\left( \frac{N}{2}-1, \frac{N}{2} \right) \\
 &- 2 E\left( \frac{N}{2}, \frac{N}{2} \right), 
\end{split}
\end{equation}
for even integer $N(\ge4)$, 
in analogy with the studies of fermionic Hubbard models. 
Here, $E(N_\ua,N_\da)$ is the ground-state energy for fixed particle numbers, $N_\ua$ and $N_\da$. 
The results for the $SU(2)$-symmetric case $g_{\ua\da}=g$ are presented in Fig.~\ref{fig:Gap}. 
We find upward spikes at $\nu/2=k/3$ with integer $k$, 
which indicate the appearance of gapped uncondensed states at these filling factors. 

%--------------------------------------
% [ Abelian Halperin (221) state ]

The appearance of a gap at $\nu=1/3+1/3$ can be naturally interpreted 
from the formation of the Halperin $(221)$ state \cite{Halperin83,Paredes02}, which is an Abelian spin-singlet state. 
On a disc geometry, its wave function is written as 
\begin{equation}\label{eq:221}
 \Psit^{221} = \prod_{i<j} (z_i^\ua-z_j^\ua)^2 \prod_{i<j} (z_i^\da-z_j^\da)^2 \prod_{i,j} (z_i^\ua-z_j^\da).  
\end{equation}
Here, as in Ref.~\cite{Read99}, a tilde on the wave function indicates 
that it has to be multiplied by the usual Gaussian factors for the $xy$ plane. 
The contact interactions in Eq.~\eqref{eq:Hint} vanish for this wave function,  
and therefore Eq.~\eqref{eq:221} is an exact zero-energy ground state for arbitrary $g_{\ua\da}\ge 0$ and $g\ge 0$. 
On a torus, there appear triply-degenerate zero-energy ground states due to the center-of-mass degeneracy. 
Performing exact diagonalization up to $N_V=15$ and $N_\ua=N_\da=5$ with $g_{\ua\da}=g$, 
we find a stable energy gap of magnitude $\sim 0.035 g$ above these zero-energy states. 

%--------------------------------------
% [Non-Abelian spin-singlet state]

We now investigate the origins of the gapped states at $\nu=k/3+k/3$ with $k\ge 2$. 
Precisely at these filling factors, Ardonne and Schoutens \cite{Ardonne99} proposed 
non-Abelian extensions of the Halperin $(221)$ state, termed the $SU(3)_k$ states \cite{Comment_NASS}. 
On a disc, their wave functions are written as 
\begin{equation}\label{eq:SU3_k}
 \Psit^{SU(3)}_k = \Scal_\group \prod_\group \Psit^{221}.
\end{equation}
In this construction, the $N=rk+rk$ bosons are first partitioned into $k$ groups, 
each with $r$ particles in each spin state $\ua$, $\da$. 
For each group we write a Halperin $\Psit^{221}$ factor, 
and then such factors are multiplied together. 
Finally, we apply the symmetrization operation $\Scal_\group$ over all different ways of dividing the particles into $k$ groups. 
Our motivation to consider the states \eqref{eq:SU3_k} stems from the analogy with the scalar case; 
their single-component counterparts, Read-Reazayi [$SU(2)_k$] states \cite{Read99},  
give good approximations to the ground states of the scalar Bose gas \cite{Cooper01}.
The wave function \eqref{eq:SU3_k} is a unique zero-energy eigenstate of a Hamiltonian 
consisting of a $(k+1)$-body interaction \cite{Ardonne01}: 
\begin{equation} \label{eq:H_k}
 H_k = \sum_{i_1<\cdots<i_{k+1}} \delta (z_{i_1}-z_{i_2}) \cdots \delta (z_{i_k}-z_{i_{k+1}}), 
\end{equation}
where the indices $i_1,\dots,i_{k+1}$ run over all the particles of both components. 
On a torus, this Hamiltonian leads to a ground-state degeneracy of 
$(k+1)(k+2)/2$ \cite{Ardonne01}. 
The appearance of this non-trivial degeneracy is related to an underlying topological order, 
and can be used as an indicator of the corresponding topological phase. 

%############################
\begin{figure}
\begin{center}
\includegraphics[width=0.47\textwidth]{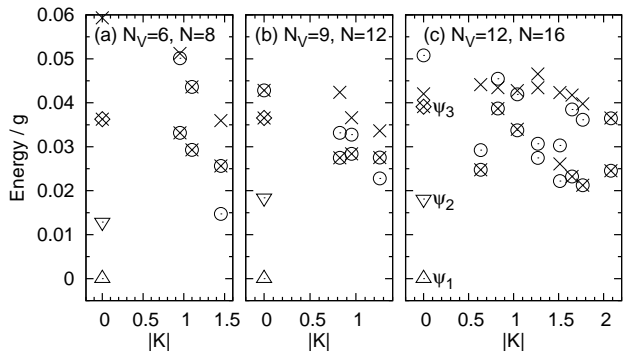}
\cutspace
\end{center}
\caption{
Energy spectra versus the pseudomomentum $\Kv$ 
at the filling factor $\nu=2/3+2/3$ for the $SU(2)$-symmetric case $g_{\ua\da}=g$. 
The ground-state energy is subtracted from the spectrum. 
For $N_V=6$ and $9$, the same aspect ratios as in Fig.~\ref{fig:Gap} are used; 
for $N_V=12$, we set $L_x/L_y=4/(3\sqrt{3})$. 
Upward ($\triangle$) and downward ($\triangledown$) triangles, representing $\psi_1$ and $\psi_2$ respectively, 
indicate the two lowest-energy states in the sector with $\Kv=0$, $\rho_\pi=+1$, and $p_{\ua\da}=+1$.   
Diamonds, representing $\psi_3$, indicate the lowest-energy state in the sector with $\Kv=0$, $\rho_\pi=-1$, and $p_{\ua\da}=-1$. 
Circles indicate other eigenstates in the equal-population case $N_\ua=N_\da=N/2$. 
Crosses show eigenstates for the minimally imbalanced case $N_\ua=N/2+1$ and $N_\da=N/2-1$. 
Only the two lowest energies are displayed in each sector. 
}
\label{fig:ener_K}
\end{figure}
%############################

%--------------------------------------
% [ Topological degeneracy ]

To discuss the possibility of $SU(3)_2$ state at $\nu=2/3+2/3$, 
we present the energy spectra versus the pseudomomentum $\Kv$ in Fig.~\ref{fig:ener_K}. 
At $\Kv=0$, we further decompose the Hilbert space 
using quantum numbers $\rho_\pi=\pm 1$ and $p_{\ua\da}=\pm 1$ 
for the $\pi$ spatial rotation $R_\pi$ and the interchange of two components, $P_{\ua\da}$, respectively. 
In addition to the equal-population case $N_\ua=N_\da=N/2$, 
we also present data for the minimally imbalanced case $N_\ua=N/2+1$ and $N_\da=N/2-1$ by cross symbols 
for $N=8$ and $12$; 
equal-population states which are not degenerate with crosses are identified as spin singlets. 
We find that for all the system sizes examined in Fig.~\ref{fig:ener_K}, 
the two lowest-energy states are spin singlets in the sector with $\Kv=0$, $\rho_\pi=+1$ and $p_{\ua\da}=+1$ (labeled as $\psi_1$ and $\psi_2$).  
By exact diagonalization of the $3$-body Hamiltonian [Eq.~\eqref{eq:H_k} with $k=2$] on a torus, 
we find doubly-degenerate zero-energy eigenstates in the same sector. 
Therefore, $2$-body and $3$-body interactions seem to show a consistent topological degeneracy; 
due to a center-of-mass degeneracy of $3$, they both lead to the total degeneracy of $6$.  
Although the splitting between $\psi_1$ and $\psi_2$  and the gap above them 
still show irregular dependences on $N$ in Fig.~\ref{fig:ener_K}, 
the consistency of the quantum numbers of these lowest-energy states 
with those of the $3$-body Hamiltonian provides an evidence 
for the formation of $SU(3)_2$ states.   

%############################
\begin{figure}
\begin{center}
\includegraphics[width=0.50\textwidth]{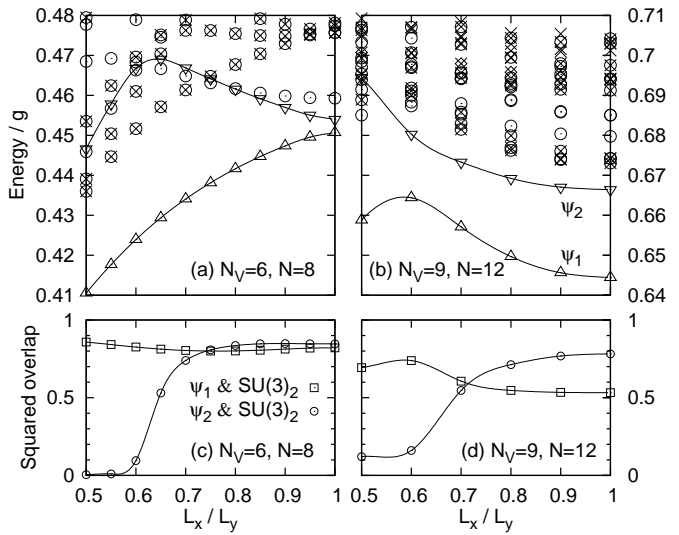}
\cutspace
\end{center}
\caption{
(a,b) Energy spectra versus the aspect ratio $L_x/L_y$, 
at the filling factor $\nu=2/3+2/3$ in the $SU(2)$-symmetric case $g_{\ua\da}=g$. 
The same symbols as in Fig.~\ref{fig:ener_K} are used. 
Solid curves are guides for the eyes for the states $\psi_{1,2}$.  
(c,d) Squared overlaps of $\psi_{1,2}$ with the $SU(3)_2$ states, plotted against the aspect ratio. 
}
\label{fig:ener_Lx}
\end{figure}
%############################

%--------------------------------------
% Aspect ratio dependence

In Fig.~\ref{fig:ener_Lx}, we examine the dependence on the aspect ratio $L_x/L_y$ 
(only the case of $L_x/L_y\le 1$ is examined without loss of generality). 
The $\psi_1$ and $\psi_2$ states continue to be the two lowest-energy states 
in the ranges $0.8 \lesssim L_x/L_y \le 1$ and $0.6 \lesssim L_x/L_y \le 1$ for $N=8$ and $12$, respectively  
[Fig.~\ref{fig:ener_Lx}(a,b)].  
Such robustness of the spectral structures under a change in the ratio $L_x/L_y$ 
is generally expected in vortex liquid states, 
which do not break the translational symmetry. 
We also calculated the overlaps of $\psi_1$ and $\psi_2$ with the $SU(3)_2$ states 
obtained as the ground states of Eq.~\eqref{eq:H_k} 
[Fig.~\ref{fig:ener_Lx}(c,d)].   
The average of the squared overlaps is around $0.8$ and $0.6$ for $N=8$ and $12$, respectively, 
in the range $0.7\lesssim L_x/L_y \le 1$ 
(for $L_x/L_y\lesssim 0.7$, one of the overlaps decreases rapidly) \cite{Comment_ovlp}. 
Although the overlaps decrease as a function of $N$, 
the increasing range of $L_x/L_y$ with the robust spectral structures 
supports the appearance of the $SU(3)_2$ state in the thermodynamic limit.

%############################
\begin{figure}
\begin{center}
\includegraphics[width=0.38\textwidth]{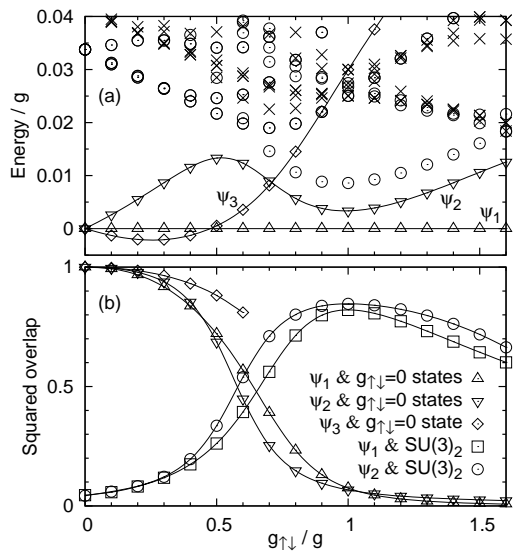}
\cutspace
\end{center}
\caption{
(a) Energy spectrum versus coupling ratio $g_{\ua\da}/g$ for $N_V=6$ and $N=8$ with $L_x/L_y=1$. 
The states $\psi_{1,2,3}$ and symbols are defined in a similar manner as in Fig.~\ref{fig:ener_K}. 
The energy of $\psi_1$ is subtracted from the whole spectrum. 
(b) Squared overlaps of the lowest-energy states $\psi_{1,2,3}$ with the states at $g_{\ua\da}=0$ and the $SU(3)_2$ states, 
plotted against $g_{\ua\da}/g$.  
}
\label{fig:ener_g}
\end{figure}
%############################

%--------------------------------------
% [ Phase transition ]

Finally, we discuss a phase transition which occurs as the ratio $g_{\ua\da}/g$ is tuned at $\nu=2/3+2/3$. 
For $g_{\ua\da}=0$, the system consists of decoupled scalar bosons;  
this system has been argued to show an incompressible composite fermion state 
at $N_\alpha/N_V=2/3$ \cite{Regnault03}. 
On a torus, each component shows a triple center-of-mass degeneracy of ground states. 
Thus, the two-component system shows $9$-fold degenerate ground states, 
each of which is given by the direct product of the above 3 states.  
In Fig.~\ref{fig:ener_g}(a), we present the energy spectrum as a function of $g_{\ua\da}/g$. 
For small $g_{\ua\da}/g$, 
we find three low-energy levels, which are separated from higher levels by a finite gap; 
each level is triply degenerate, implying the persistence of a quasi-degeneracy of $9$ in this regime. 
For larger $g_{\ua\da}/g (\gtrsim 0.6)$, we find that the energy of $\psi_3$ goes up, while $\psi_1$ and $\psi_2$ remain the lowest two states, 
leading to a total quasi-degeneracy of $6$ as expected for the $SU(3)_2$ state. 
These results point to a phase transition 
from the direct product of composite fermion states 
to the $SU(3)_2$ state at $g_{\ua\da}/g\approx 0.6$. 
Essentially the same behavior is found for a larger system size ($N_V=9$ and $N=12$; not shown), 
although the quasi-degeneracy of $\psi_1$ and $\psi_2$ around $g_{\ua\da}/g=1$ is less evident than in Fig.~\ref{fig:ener_g}(a). 

The occurrence of the phase transition can be further supported 
by the results of the squared overlaps in Fig.~\ref{fig:ener_g}(b). 
For small $g_{\ua\da}/g$, 
the states $\psi_{1,2,3}$ continue to have large overlaps with those in the decoupled case $g_{\ua\da}=0$. 
For $g_{\ua\da}/g (\gtrsim 0.6)$, $\psi_{1,2}$ have larger overlaps with the $SU(3)_2$ states 
than with the states at $g_{\ua\da}=0$. 
It remains unclear whether a phase transition occurs directly between the two quantum Hall states 
or any intermediate phase exists. 

%--------------------------------------
% [ Conclusion ]

In summary, we have studied quantum Hall states in rapidly rotating two-component Bose gases. 
We have presented numerical evidences that a NASS state appears at $\nu=2/3+2/3$. 
Non-Abelian anyonic excitations in the NASS state can carry spins \cite{Ardonne99,Ardonne01}, 
and spin-selective operations would potentially offer better probe and control of such excitations 
than in scalar Bose gases. 
% Since the NASS state features non-Abelian statistics of quasi-hole excitations \cite{Ardonne99,Ardonne01}, 
% the present result opens up a possibility of observing novel quasi-particle statistics in two-component gases. 
Furthermore, we have demonstrated that changing the ratio $g_{\ua\da}/g$ drives a phase transition 
from the product of separate quantum Hall states to a spin-singlet quantum Hall state. 
Systematic appearance of spikes at $\nu=k/3+k/3$ with integer $k$ in Fig.~\ref{fig:Gap} 
suggests that $SU(3)_k$ states with $k\ge 3$ might also appear 
for the realistic contact interactions \eqref{eq:Hint}. 
More precise characterizations of the ground states at these filling factors 
deserve further studies. 

%--------------------------------------
% [ Acknowledgement ]

This work was supported by 
a Grant-in-Aid for Scientific Research on Innovative Areas ``Topological Quantum Phenomena'' (No. 22103005), 
KAKENHI 22340114, 
a Global COE Program ``the Physical Science Frontier'', 
and the Photon Frontier Network Program, 
from MEXT of Japan. 

%--------------------------------------
% [ Comment ]

{\it Note added.} 
During the preparation of this Rapid Communication, 
we became aware of an independent work by Gra{\ss} {\it et al.} \cite{Grass12}, 
where similar numerical evidences for the NASS states are presented. 

%During the preparation of this Letter, 
%a preprint by Gra{\ss} {\it et al.} \cite{Grass12} appeared 
%which presents similar numerical evidences for the NASS states. 

\cutspace

%############################
% \begin{figure}
% \begin{center}
% \includegraphics[width=0.50\textwidth]{.eps}
% \end{center}
% \caption{
% }
% \label{fig:}
% \end{figure}
%############################

%%%%%%%%%%%%%%%%%%%%%%%%%%%%%%%%%%%%%%%%%%%%%%%%%
%References
%%%%%%%%%%%%%%%%%%%%%%%%%%%%%%%%%%%%%%%%%%%%%%%%%
\newcommand{\etal}{{\it et al.}}
\newcommand{\PRL}[3]{Phys. Rev. Lett. {\bf #1}, \href{http://link.aps.org/abstract/PRL/v#1/e#2}{#2} (#3)}
\newcommand{\PRLp}[3]{Phys. Rev. Lett. {\bf #1}, \href{http://link.aps.org/abstract/PRL/v#1/p#2}{#2} (#3)}
\newcommand{\PRA}[3]{Phys. Rev. A {\bf #1}, \href{http://link.aps.org/abstract/PRA/v#1/e#2}{#2} (#3)}
\newcommand{\PRAp}[3]{Phys. Rev. A {\bf #1}, \href{http://link.aps.org/abstract/PRA/v#1/p#2}{#2} (#3)}
\newcommand{\PRAR}[3]{Phys. Rev. A {\bf #1}, \href{http://link.aps.org/abstract/PRA/v#1/e#2}{#2} (R) (#3)}
\newcommand{\PRB}[3]{Phys. Rev. B {\bf #1}, \href{http://link.aps.org/abstract/PRB/v#1/e#2}{#2} (#3)}
\newcommand{\PRBp}[3]{Phys. Rev. B {\bf #1}, \href{http://link.aps.org/abstract/PRB/v#1/p#2}{#2} (#3)}
\newcommand{\PRBR}[3]{Phys. Rev. B {\bf #1}, \href{http://link.aps.org/abstract/PRB/v#1/e#2}{#2} (R) (#3)}
\newcommand{\arXiv}[1]{arXiv:\href{http://arxiv.org/abs/#1}{#1}}
\newcommand{\condmat}[1]{cond-mat/\href{http://arxiv.org/abs/cond-mat/#1}{#1}}
\newcommand{\JPSJ}[3]{J. Phys. Soc. Jpn. {\bf #1}, \href{http://jpsj.ipap.jp/link?JPSJ/#1/#2/}{#2} (#3)}
\newcommand{\PTPS}[3]{Prog. Theor. Phys. Suppl. {\bf #1}, \href{http://ptp.ipap.jp/link?PTPS/#1/#2/}{#2} (#3)}
\newcommand{\hreflink}[1]{\href{#1}{#1}}


\begin{thebibliography}{99}
\bibitem{Comment_contact}
 These are obtained by multiplying the factor $\sqrt{M\omega_\parallel/(2\pi \hbar)}$ to the coupling constants 
 $g_\alpha^\ThD=4\pi\hbar^2 a_\alpha/M$ and $g_{\ua\da}^\ThD=4\pi\hbar^2 a_{\ua\da}/M$ 
 for the three-dimensional contact interactions. 
 This factor arises from the restriction to the ground state of the axial confinement.
\bibitem{Comment_NASS} 
 This is the $n=2$ case of $SU(n+1)_k$ states 
 proposed for $n$-component systems in Ref.~\cite{Reijnders02}.
\bibitem{Comment_ovlp}
 For $N=4$, the overlap is perfect (exactly $1$) for any aspect ratio $L_x/L_y$, 
 due to a limited Hilbert space.
\end{thebibliography}
\end{document}